\documentclass[11pt]{article}
\usepackage{epsfig,sint,macros}
\begin{document}
\begin{titlepage}
\begin{flushright}
  DESY-98-064\\
  OUTP-98-45-P
\end{flushright}

\vskip 1 cm
\begin{center}
  {\Large\bf Precision computation of a low-energy reference scale\\[0.5ex]
  in quenched lattice QCD }
\end{center}
\vskip 1 cm
\vbox{
\centerline{
\epsfxsize=2.5 true cm
\epsfbox{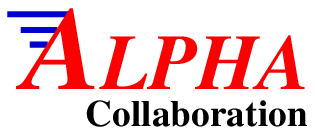}}
}
\vskip 1 cm
\begin{center}
{\large Marco Guagnelli$^{\scriptscriptstyle a}$, 
          Rainer Sommer$^{\scriptscriptstyle a}$ and
          Hartmut Wittig$^{\scriptscriptstyle b}$}
\vskip 1.0cm
$^{\scriptstyle a}$
DESY-Zeuthen \\
Platanenallee 6, D-15738 Zeuthen
\vskip 2.5ex
$^{\scriptstyle b}$
Theoretical Physics, University of Oxford \\
1~Keble Road, Oxford OX1~3NP, UK
\vskip 1.5cm
{\bf Abstract}
\vskip 0.7ex
\end{center}
We present results for the reference scale $r_0$ in SU(3) Lattice
Gauge Theory for $\beta=6/g_0^2$ in the range
$5.7\leq\beta\leq6.57$. The high relative accuracy of 0.3--0.6\% in
$r_0/a$ was achieved through good statistics, the application of a
multi-hit procedure and a variational approach in the computation of
Wilson loops. A precise definition of the force used to extract $r_0$
has been used throughout the calculation which guarantees that $r_0/a$
is a smooth function of the bare coupling and that subsequent
continuum extrapolations are possible. The results are applied to the
continuum extrapolations of the energy gap $\Delta$ in the static
quark potential and the scale $\Lmax/r_0$ used in the calculation of
the running coupling constant.  \vfill

\begin{flushleft}
  DESY-98-064 \\
  OUTP-98-45-P \\
 June 1998
\end{flushleft} 

\eject

\vfill

\eject

\end{titlepage}

\section{Introduction \label{s_intro}}
Monte Carlo calculations of lattice QCD in the quenched approximation
and in the SU(3) Yang-Mills theory have reached considerable
precision. For example, the accuracy of hadron mass calculations using
the Wilson action is quoted to be well below 1\% at finite values of
the lattice spacing \cite{hadr:CPPACS}, and the running of the
coupling in a specific non-perturbative scheme is known to a precision
of around $1\%$ over energy scales varying by two orders of magnitude
\cite{alpha:SU3,lat97:martin}. In the latter case, the continuum
limit was taken.

It was already noted in \cite{pot:r0_SCRI} that in comparison to such
a level of accuracy, a low-energy reference scale in the Yang-Mills
theory is known with much worse precision, despite the fact that such
a scale is very important for the analysis of the results. In the
second example of the running coupling, a low-energy reference scale
is required in order to determine the overall momentum scale for which
the coupling was computed (see \sect{s_lmax}). In the first example, a
gluonic reference scale is very useful in order to assess the size of
lattice artifacts present in different discretizations of QCD
\cite{impr:SCRI} and to provide a check on the continuum
extrapolations presented in \cite{hadr:CPPACS}, which have been
performed using hadronic scales. The reason is that, in the quenched
approximation, the leading cutoff effects in gluonic observables are
proportional to $a^2$, only. For instance, for Wilson fermions without
improvement, the ratio of a hadronic and a gluonic mass scale
approaches the continuum limit with an asymptotic rate proportional to
the lattice spacing $a$, where the linear $a$-dependence originates
purely from the hadron mass.  Obviously, good precision in the gluonic
reference scale is desirable for such applications.

In \cite{pot:r0} a reference scale, $r_0$, was introduced for such
purposes. This length scale is defined in terms of the force, $F(r)$,
between external static charges in the fundamental representation. It
is the solution of
\bes
  r_0^2 F(r_0) = 1.65 \, . \label{e_r0_cont}
\ees
The constant on the right hand side was chosen such that $r_0$ has a
value of approximately $0.5\,\fm$ in QCD \cite{pot:r0}. In comparison
to glueball masses and the string tension, it is much easier to
compute with controlled errors \footnote{The reason is that glueball
correlation functions are quite noisy; the string tension refers to
the asymptotic force at infinite distance, which is difficult to
extract without additional assumptions about the leading corrections
to the large-distance behaviour.}. 
 
However, for gauge group SU(3) and the Wilson plaquette action, $r_0$
has been computed only with modest precision so far. A recent new
effort \cite{pot:r0_SCRI} concentrated on the region of small $\beta$
/ large lattice spacing.  In this paper we compute $r_0$ down to small
lattice spacings $a\approx 0.04\,\fm$, covering the whole range
$0.17\,\fm\,\,\lesssim\,\, a\,\,\lesssim\,\,0.04\,\fm$ and improving
the precision by a factor up to five.  This gain in precision is only
to a small part due to the use of a modern parallel computer (APE100),
but results mainly from the application of the known methods of
variance reduction \cite{PPR} and a variational calculation
\cite{varia:michael,phaseshifts:LW}, as employed already in
\cite{pot:r0} for gauge group SU(2).

Our results are given in the form of a table and also an interpolating
fit function which describes $r_0$ with a precision between $0.3\%$
(at $\beta\approx 5.7$) to $0.6\%$ (at $\beta\approx 6.57$). At the
level of such precision, one can expect that the results depend on the
chosen discretization of the force $F(r)$. One and the same definition
has to be used at all values of $\beta$ in order to guarantee a smooth
approach to the continuum limit. The procedure applied in
\cite{pot:r0_SCRI} and \cite{lat97:bali} does not correspond to a
precise definition of the force satisfying this criterion. Indeed, our
results differ slightly from the ones reported in
\cite{pot:r0_SCRI,lat97:bali}. 
The differences are only $2.3\pm0.9$\% at $\beta=5.7$,
$0.9\pm0.5$\% at $\beta=5.85$ and
$-1.2\pm0.6$\% at $\beta=6.2$.
From our study of finite size
effects we can exclude that these differences are due to different
lattice sizes used in the various calculations.

We apply our results for $r_0$ to the calculation and continuum
extrapolation of the ratio $\Lmax/r_0$, which determines the scale in
the running coupling and $\Lambda$-parameter calculations of
\cite{lat97:martin}. We also compute the gap between the 
ground state potential and its first excitation in the continuum limit.
 
This paper is organized as follows. In the following section we
describe the computation of $r_0$ in lattice units, starting with a
recapitulation of the precise definition. Readers uninterested in
these details will find the results in \sect{s_r0scale} and the
applications to two continuum extrapolations in \sect{s_cont}. We
finish with a brief conclusion.

\section{Determination of $r_0$ \label{s_scale}}
\subsection{Definition of the force \label{s_def}}
The definition \eq{e_r0_cont} is unique in the continuum. For finite
lattice spacing, we have to specify which discretization of the force
is to be used. Lattice artifacts -- and therefore the precise values
of $r_0/a$ -- do depend on these details. However, it has become
customary to replace the precise definition \cite{pot:r0} by a fitting
procedure, with fits performed to potential values in the
neighbourhood of $r_0$ \cite{pot:bali_last,lat97:bali,pot:r0_SCRI}.
The motivation for such a procedure is to increase the statistical
precision, but then it is not guaranteed that $r_0/a$ is a smooth
function of $\beta$, allowing for systematic continuum
extrapolations. We will therefore return to a precise definition and
demonstrate that the achieved statistical precision is in fact very
good.

We start from potential values, $V(\vecr)$, given along one fixed
orientation $\vecd/|\vecd|$ on the lattice. Here, $\vecd$ is a
(three-dimensional) vector on the lattice, and $\vecr$ is an integer
multiple of $\vecd$. For example we may have $\vecd=(a,a,0)$. The
force is then defined as
\bes
 F_{\vecd}(\rI) &=& {V(\vecr)  - V(\vecr-\vecd) 
                         \over 
                          |\vecd|     } \\
        {1 \over 4 \pi \rI^2}  &=&  - {G(\vecr)  - G(\vecr-\vecd) 
                         \over 
                          |\vecd|  } \\
  G(\vecr)  & = & a^{-1} \int_{-\pi}^{\pi} 
                 {\rmd^3 k \over (2\pi)^3}
                 {\prod_{j=1}^3 \cos(r_jk_j/a)   \over 
                          4  \sum_{j=1}^3 \sin^2(k_j/2)   } \, .                                   
\ees
The particular choice of $\rI$ ensures that the force is given by
$F_{\vecd}(\rI)=\frac43 g^2 /(4 \pi \rI^2) +\rmO(g^4)$. To lowest
order of perturbation theory, it eliminates the lattice artifacts
exactly; they remain (probably quantitatively reduced) only in the
higher $\rmO(g^4)$ terms. The force $F_{\vecd}(\rI)$ as defined
above is therefore called a tree-level improved observable.

In the following, we choose $\vecd=(a,0,0)$ dropping this index on the
force. This choice has also been made in \cite{pot:r0_UKQCD} and our
data may therefore be directly compared with the results from that
reference. Solving \eq{e_r0_cont} requires furthermore an
interpolation of $F(\rI)$, which is easily done with small systematic
errors \cite{pot:r0}.

If the procedure to compute the scale $r_0$, as outlined in this
subsection, is applied consistently at all values of $\beta$
considered, then $r_0/a$ will be a smooth function of the bare
coupling.

\subsection{Wilson loop correlation matrix}
We now turn to the details of the computation of $V(r)$. It is well
known that an effective calculation of the potential starts from
smeared Wilson loops. Our smearing operator $\smear$ acts on the
spatial components of the gauge fields via \cite{smear:ape}
\bes
 \smear U(x,k)& =& 
   \proj \{\,
           U(x,k) \,+\, \alpha \, \sum_{j\ne k} \, [ \,
           U(x,j)U(x+a\hat{j},k)U^{\dagger}(x+a\hat{k},j) + \nonumber
           \\
           && \qquad \qquad \quad U^{\dagger}(x-a\hat{j},j)U(x-a\hat{j},k)
                         U(x+a\hat{k}-a\hat{j},j)\, ]
         \,\} \, ,
\ees
where $\proj$ denotes a projection back into the group SU(3) and the
SU(3) gauge field residing on a link from $x$ to $x+a\hat{\mu}$ is
denoted by $U(x,\mu)$. For different smearing levels $l=0, \ldots,
M-1$ we then construct smeared spatial links according to
\bes
        U_l(x,k) &=&  \smear^{n_l} U(x,k) \,.
\ees
For the time-like links in the Wilson loops, we apply the 
multi-hit procedure \cite{PPR} in order to reduce the variance.
These multi-hit averaged links\footnote{
We applied an average using 10 Cabibbo-Marinari 
iterations each of which consisted of updates in three SU(2) 
subgroups embedded in SU(3).} 
are denoted by $\Ubar$. At fixed $r$, an $M\times M$ correlation
matrix of Wilson loops is then formed as
\bes
C_{lm}(t) = \left\langle \tr\left\{ V_l(0,r\hat{1})
            \Vbar(r\hat{1},r\hat{1}+t\hat{0}) 
            V_m^{\dagger}(t\hat{0},r\hat{1}+t\hat{0})
            \Vbar^{\dagger}(0,t\hat{0})\right\} 
            \right\rangle = C_{ml}(t) \,,
\ees
where
\bes
 V_l(x,x+r\hat{1}) &=& U_l(x,1) U_l(x+a\hat{1},1)  \ldots  
                  U_l(x+(r-a)\hat{1},1), \\
 \Vbar(x,x+t\hat{0}) &=&  \Ubar(x,0) \Ubar(x+a\hat{0},0)  \ldots  
                  \Ubar(x+(t-a)\hat{0},0)\,.    
\ees 
\begin{table}[htb] 
\centering
\begin{tabular}{ r  l  r  c  cc r l r c}
\hline \\[-1.0ex]
$L/a$ & $\beta$ & $N_{\rm meas}$ & $M$ &&& 
$L/a$ & $\beta$ & $N_{\rm meas}$ & $M$  \\[1.0ex] 
\hline \\[-1.0ex]
 10  & 5.7    & $1000$  & $4$ &&& 24  & 6.2    &  $400$    & $4$ \\
 12  & 5.8    & $3200$  & $4$ &&& 32  & 6.4    &  $200$    & $3$ \\
 16  & 5.95   &  $600$  & $4$ &&& 40  & 6.57   &  $572$    & $1$ \\
 20  & 6.07   &  $780$  & $4$ &&&     &        &           &
\\[1.0ex]
\hline 
\end{tabular} 
\caption[t_param]{\footnotesize Simulation parameters. $N_{\rm meas}$
denotes the number of different gauge field configurations for which
the correlation matrices were computed. For every configuration
averages were computed over all $(L/a)^4$ points of the lattice and
three permutations of the lattice axes. $M$ is the number of smearing
levels. \label{t_param}}
\end{table} 

These correlation matrices were computed on lattices generated by a
hybrid over-relaxation algorithm \cite{HOR1,HOR2} with a mixture of
$N_{\rm or} \approx 1.5 \, \rnod/a$ over-relaxation steps per heat
bath step. In each simulation, a measurement of Wilson loops is
performed every $20\times(N_{\rm or}+1)$ sweeps. The usual
binning--jackknife procedure was used to compute the errors, and to
control autocorrelations in the measurements. We further chose
\bes
  \alpha={1 \over 2}\,,\quad
  n_l \approx {l\over2}\left({r_0}\over{a}\right)^2\,.
\label{e_alpha}
\ees
The value we used for $n_2$ corresponds roughly to what was estimated
to be the optimal smearing in \cite{pot:r0_SCRI}. Other parameters
are listed in \tab{t_param}. Note that for $\beta=6.57$ memory
limitations restricted us to one smearing level ($M=1$), and deviating
from \eq{e_alpha}, we used $n_0=160$ in this case.

\subsection{Ground state potential and gap}
Following \cite{varia:michael,phaseshifts:LW}, the correlation
matrices were analyzed using a variational method: take $r$ and $t_0$
fixed ($t_0=a$ in practice) and solve the generalized eigenvalue problem,
\bes
 C(t) v_{\alpha}(t) = 
        \lambda_{\alpha}(t) C(t_0) v_{\alpha}(t) \,,
\label{e_eigenvalue_problem}
\ees
for the eigenvalues $\lambda_{\alpha}(t)$. These are related to the
different potential levels by \cite{phaseshifts:LW}
\bes
a V_{\alpha} = \ln(\lambda_{\alpha}(t) /\lambda_{\alpha}(t+a)) +
            \rmO(\rme^{-(V_M - V_{\alpha})t}) \,. \label{e_valpha}
\ees
Here, the potential levels $V_{\alpha}$ are the eigenvalues of the
(lattice) Hamilton operator in the sector of the Hilbert space that
has the quantum numbers of the usual central potential $V(r) \equiv
V_0$. Besides the potential, we are also interested in the gap 
$\Delta = V_1 - V_0$, i.e. the energy difference between the ground
state and the first excited state, since this gap controls the finite
time corrections in the extraction of $V(r)$ (see below).
 
From \eq{e_valpha} we infer that $\Delta$ can be estimated by
\bes
a \Delta  = \ln(\lambda_{1}(t) /\lambda_{1}(t+a)) -
\ln(\lambda_{0}(t) /\lambda_{0}(t+a))  + \rmO(\rme^{-(V_M - V_{2})t}) \,.
      \label{e_Delta}
\ees 
Checking our data for convergence at large $t$, we observed that the
systematic corrections in \eq{e_Delta} are smaller than the
statistical errors for $t \geq \frac32 r_0$. This value of $t$ was
then used for our estimates of $\Delta$. We will return to these
results in \sect{s_gap}. Here we only note that $\Delta|_{r=\rnod}
\approx 3.3 /\rnod$ in the continuum limit.

\begin{figure}[ht]
\hspace{0cm}
\vspace{-0.0cm}

\centerline{
\psfig{file=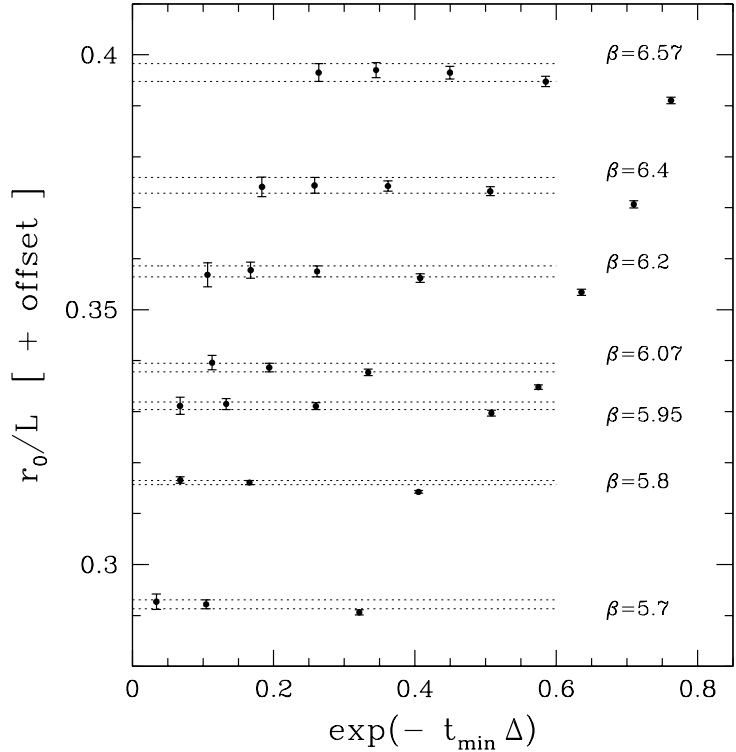,width=10cm}
}
\vspace{-0.0cm}
\caption{\footnotesize
Estimates $r_0/a$ for various minimum time separations, $\tmin$, in
the fits \eq{e_potfits}.  $\beta$ increases from bottom to top and the
values for different $\beta$ have been shifted relative to each other
for clarity. The $x$-coordinate corresponds to the slowest varying
finite-$t$ correction term (as a function of $\tmin$). For the gap
$\Delta$ the value $\Delta \approx 3.3/r_0$ has been used. Dashed
error bands denote our final estimates.
\label{f_r0_textrap}}
\end{figure}

As a first step in our analysis, the corrections to the ground state
potential $V(r)$ itself were studied using \eq{e_valpha} and found to
be much smaller than in the case of the gap $\Delta$. However, we were
not satisfied with the statistical errors of $\ln(\lambda_{0}(t)
/\lambda_{0}(t+a))$. We therefore considered an alternative way of
extracting $V(r)$ by forming the projected Wilson loop correlation
\bes
 W(t) = v_0^{\rm T} C(t) v_0 \,,
\ees
where $v_0$ is the eigenvector computed for $t=t_0+a$ in
\eq{e_eigenvalue_problem}. The projected correlation $W(t)$ then turns
out to be strongly dominated by the ground state. Next, $W(t)$ was
fitted to a single exponential
\bes
 W(t) \approx A \rme^{-V(r)t} \label{e_potfits}
\ees
for $t$ ranging from $\tmin$ to $\tmax$, usually setting
$\tmax=\tmin+3a$. The parameters in these fits converge to the true
values with corrections of order $\exp(-\tmin\Delta)$. To demonstrate
that these corrections are under control, we turn directly to the
desired quantity $r_0$. Using $V(r)$ obtained from fits with fixed
$\tmin$, we computed $r_0$ as outlined in \sect{s_def}. We show in
\fig{f_r0_textrap} the values of $r_0$ as a function of
$\exp(-\tmin\Delta)$. Convergence is observed for large values of
$\tmin$. As our final values we took the fits with $\tmin$ such that
$\exp(-\tmin\Delta) < 0.3$. Let us summarize some important points
about this analysis.
\vspace{-0.5ex}
\begin{itemize}
\itemsep 0 pt
\item{Like in the calculation of any energy value from Euclidean 
      correlation functions at finite $t$, it is important to have an
      estimate for the gap in the channel considered. This is because
      time separations of the order of the inverse gap have to be
      reached, in order to judge whether correction terms are
      significant;}\vspace{-0.5ex}
\item{Due to the use of the variationally determined ``wavefunction'' $v_0$,
      the corrections are most likely dominated by higher states and
      decay faster than $\exp(-t\Delta)$ for the accessible values
      of~$t$. Some evidence for this is seen in \fig{f_r0_textrap}, in
      particular at larger values of $\beta$;}\vspace{-0.5ex}
\item{In most cases the fits are statistically acceptable already for 
      smaller $\tmin$ than that used for our final
      estimates;}\vspace{-0.5ex} 
\item{Our estimates obtained directly from the application of \eq{e_valpha}
      for large $t$ (starting from $t=3a$) 
      agree with our final estimates shown above;} 
      \vspace{-0.5ex}          
\item{The variance reduction \cite{PPR} is essential to obtain estimates
      with moderate errors at large $t$.} \vspace{-0.5ex}
\end{itemize}\vspace{-0.5ex}
These observations make us confident that our estimates for the
potential and $r_0$ represent solid results.

\subsection{The scale $r_0/a$ \label{s_r0scale}}

The calculation of the force $F(\rI)$ from the ground state potential
and the subsequent interpolation of the force to extract $r_0/a$ follows
the procedure outlined in \sect{s_def}.

\subsubsection{Results; finite size effects}
  \begin{table}[tb] 
  \centering
  \begin{tabular}{ l  r@{.}l  c | r@{.}l l | r@{.}l l }
  \hline
  &\multicolumn{2}{c}{\mbox{}}&&\multicolumn{2}{c}{\mbox{}}
  &&\multicolumn{2}{c}{\mbox{}}&  \\[-1.0ex]
   $\beta$     & \multicolumn{2}{c}{$\rnod/a$} & $L/\rnod$ 
               & \multicolumn{2}{c}{$\rnod/a$} & ref.
               & \multicolumn{2}{c}{$\rnod/a$} & ref.
   \\[1.0ex] 
  \hline
  &\multicolumn{2}{c}{\mbox{}}&&\multicolumn{2}{c}{\mbox{}}
  &&\multicolumn{2}{c}{\mbox{}}&  \\[-1.0ex]
  5.7  &  2&922(~9) & 3.42 &  \multicolumn{2}{c}{\mbox{}}&
&  2&990(24)$^*$ & \cite{pot:r0_SCRI} \\
  5.8  &  3&673(~5) & 3.27 &  \multicolumn{2}{c}{\mbox{}} & 
&\multicolumn{2}{c}{\mbox{}}& \\
  5.95 &  4&898(12) & 3.27 &  \multicolumn{2}{c}{\mbox{}} & 
&\multicolumn{2}{c}{\mbox{}}& \\
  6.07 &  6&033(17) & 3.32 &  \multicolumn{2}{c}{\mbox{}} & 
&\multicolumn{2}{c}{\mbox{}}& \\
  6.2  &  7&380(26) & 3.25 &  7&29(14)
                             & \cite{pot:r0_UKQCD,Wittig:1997tr}
&  7&29(4)$^*$ & \cite{lat97:bali} \\
  6.4  &  9&74~(5)  & 3.29 &  9&75(15)
                             & \cite{pot:r0_UKQCD,Wittig:1997tr}
&\multicolumn{2}{c}{\mbox{}}& \\
  6.57 & 12&38~(7)  & 3.23 &  \multicolumn{2}{c}{\mbox{}} &
&\multicolumn{2}{c}{\mbox{}}&
 \\[1.0ex]
  \hline 
  \end{tabular} 
\caption[t_param]{\footnotesize Results for $\rnod/a$. Our values and
the respective volumes on which they have been computed are shown in
columns~2 and~3. In the last four columns we show a comparison to
other results found in the literature at the same $\beta$ values. An
asterisk indicates that the precise definition of $\rnod$ has not been
used. 
\label{t_r0}
 }
  \end{table} 

%
\begin{figure}[bt]
\hspace{0cm}
\vspace{-0.0cm}

\centerline{
\psfig{file=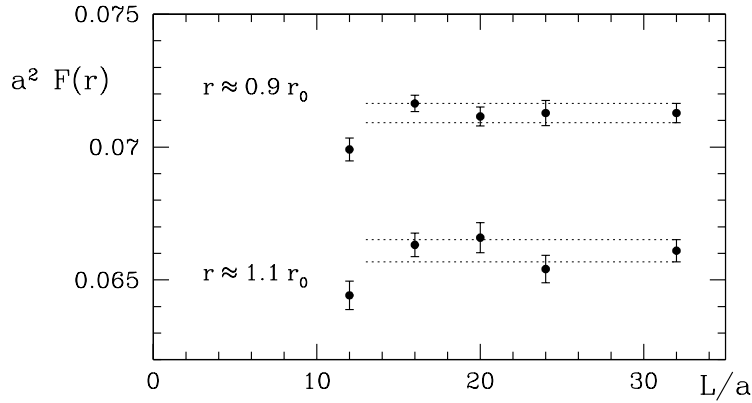,width=10cm}
}
\vspace{-0.0cm}
\caption{\footnotesize Finite size dependence of the force $F(r)$.
\label{f_L_dep}}
\end{figure}
The final values of $r_0/a$ are listed in \tab{t_r0}. We note that the
parameters $\beta$ and $L/a$ have been chosen such that $L/r_0\approx
3.3$\,, and hence the physical volume is kept constant over the whole
range of $\beta$ we considered. In order to check whether $r_0$ still
has a significant dependence on $L$ for $L\geq3.3r_0$, we carried out
a series of calculations at $\beta=5.95$ fixed but for several
different values of $L/a$. The force $F(r)$ for $r$ in the vicinity of
$r_0$ is shown in \fig{f_L_dep}. One observes that, within errors,
$F(r)$ is independent of $L$ for $L/a \geq 16$, which corresponds to
$L/r_0 \geq 3.3$. Therefore, we may take the values of $\rnod$ listed
in \tab{t_r0} as estimates in infinite volume. \Tab{t_r0} also
contains results from the literature at the same values of
$\beta$. One observes the small but significant differences quoted in
the introduction.

\subsubsection{Parametrization}
\label{s_paramet}
We now describe the parametrization of our results for $r_0/a$ in
\tab{t_r0} in terms of a smooth function of $\beta$. 
This is meant to provide an interpolating formula, so that $r_0/a$ can
be obtained at arbitrary values of $\beta$ in the interval
$5.7\leq\beta\leq6.57$.

\begin{figure}[tb]
\hspace{0cm}
\vspace{-0.0cm}

\centerline{
\psfig{file=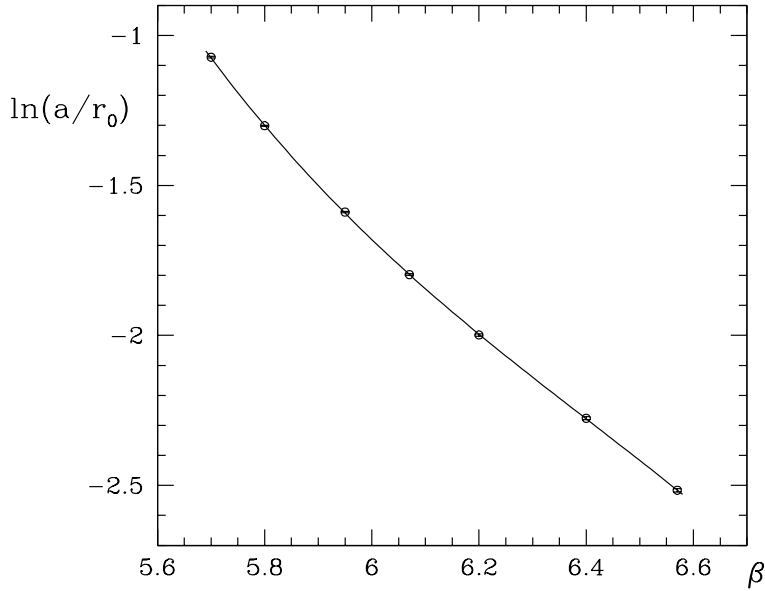,width=10cm}
}
\vspace{-0.0cm}
\caption{\footnotesize The data for $\protect\ln(a/r_0)$ (circles) and their
representation as a polynomial in $\beta$ (solid line).
\label{f_r0fit}}
\end{figure}

A convenient starting point for the parametrization is the
solution of the renormalization group equation for the bare coupling
\bes
  {a \over r_0} & = & A \,
  \rme^{-1/(2b_0{g_0^2})}\,\big(b_0g_0^2)^{-b_1/(2b_0^2)}\, 
  \rme^{-c_1 g_0^2 - c_2 g_0^4 - \ldots  },
\label{e_rg_solution}
\ees
where $b_0=11/(4\pi)^2$ and $b_1=102/(4\pi)^4$ are the universal one-
and two-loop coefficients in the perturbation expansion of the $\beta$
function, and the constant~$A$ is related to the
$\Lambda$-parameter. The contributions containing $c_1,\,c_2,\ldots$
arise from higher order terms in the perturbative $\beta$
function. From the leading behaviour we infer 
\bes
  (a/r_0) \propto \rme^{-\beta/(12b_0)}, \quad \beta=6/g_0^2,
\ees
so that we attempt a phenomenological representation of $\ln(a/r_0)$
in terms of a polynomial in $\beta$. In order to avoid large
cancellations among the fit parameters, it is advantageous to shift
the value of $\beta$, so that the intercept of the fit formula is
contained in the interval of $\beta$ values we considered. We choose
the following ansatz 
\bes
  \ln(a/r_0) = \sum_{k=0}^p\,a_k(\beta-6)^k .
\ees
A good description of our data is obtained for $p=3$:
\bes
  \ln(a/r_0) = -1.6805  -1.7139\,(\beta-6)
  +0.8155\,(\beta-6)^2 -0.6667\,(\beta-6)^3,
\label{e_poly_fit}
\ees
and a comparison of this expression with the data points is shown in
\fig{f_r0fit}. 
The deviation between the curve and the values of $r_0/a$ is 
smaller than the statistical accuracy of the data. We thus
take \eq{e_poly_fit} as our representation of $r_0/a$ in the range
$5.7\leq\beta\leq6.57$. When $r_0/a$ is evaluated using
\eq{e_poly_fit}, a relative uncertainty of 0.3\% at $\beta=5.7$ should
be assigned to the result, growing linearly to 0.6\% at
$\beta=6.57$. This level of precision roughly corresponds to the
statistical uncertainty in the data points.  All results found in the
literature, which use precisely our definition of $r_0$
\cite{pot:ukqcd65,alpha:SU3,pot:r0_UKQCD,Wittig:1997tr} are described
by \eq{e_poly_fit} within their statistical accuracy.

Another functional form of the parametrization is given directly 
by \eq{e_rg_solution}. Since the three-loop contribution $c_1$ has
been calculated in \cite{threeloop:alles}, the fit parameters are
$A,\,c_2,\,c_3,\ldots$. In order to represent the data at the level of
precision quoted above, four fit parameters are needed as in the
phenomenological fit \eq{e_poly_fit}. 
The fitted values of the parameters $c_i$ turn out to be too large to
allow for their interpretation as (effective) perturbative
coefficients -- as was expected from the well-known failure of
``asymptotic scaling'' in the bare coupling. We conclude that
\eq{e_rg_solution} does not lead to a superior representation of the
data.

\section{Continuum extrapolations \label{s_cont}}
In this section we want to give two examples of continuum
extrapolatations with the scale set by $\rnod$. The first one is the
gap $\Delta$. To our knowledge this is the first time the location of
an excited state is computed in the continuum limit of the
$4$-dimensional Yang-Mills theory.
 
\subsection{The potential gap \label{s_gap}}
%
\begin{figure}[tb]
\hspace{0cm}
\vspace{-0.0cm}

\centerline{
\psfig{file=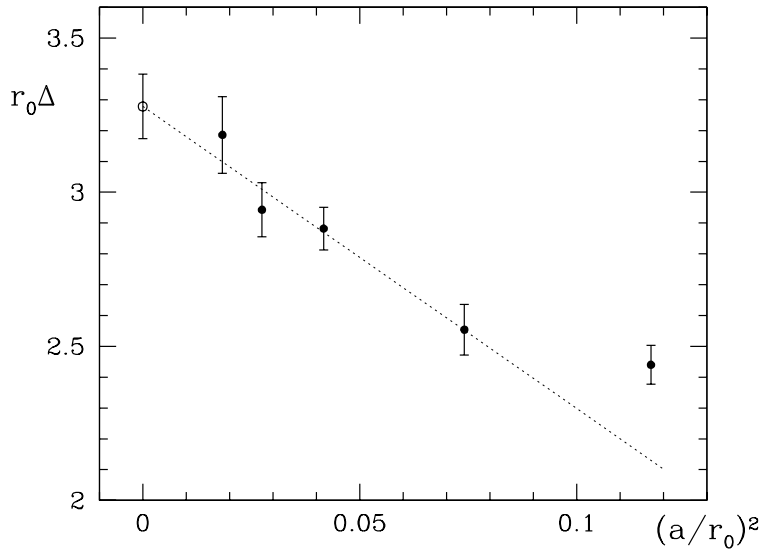,width=10cm}
}
\vspace{-0.0cm}
\caption{\footnotesize
 Continuum extrapolation of $r_0\,\Delta$.
\label{f_delta_r0}}
\end{figure}
%
The gap $\Delta$ was computed as a function of $r$ as described in the
previous section. Reliable estimates could be obtained for $\beta \leq
6.2$ only, since for the larger lattices the basis of operators
(i.e. the parameter $M$) was too small. For illustration, we
concentrate here on $\Delta|_{r=r_0}$, noting that for smaller $r$ the
gap increases, while for much larger distances it becomes difficult to
determine. To obtain the gap at $r=r_0$, an interpolation in $r$ must
be performed, which can be done easily since the dependence on $r$ is
rather weak.  We then computed the dimensionless combination $\rnod
\Delta$. The errors of this quantity are much larger than for $\rnod$
alone, namely around 2--5\%. Nevertheless, as shown in
\fig{f_delta_r0}, lattice artifacts are clearly visible. They amount
to more than 20\% at $\beta=5.8$ ($a/\rnod=0.27$) and decrease to
around 12\% at $\beta=5.95$ ($a/\rnod=0.20$).

A continuum extrapolation can be performed using a linear fit in the
leading correction term $(a/\rnod)^2$. Our graph shows the result
obtained after discarding the data point at the largest value of the
lattice spacing. The extrapolation yields 
\bes
   \rnod \Delta|_{r=\rnod} = 3.3(1)
\label{e_r0delta}
\ees
in the continuum limit, and a similar number (with larger error) is
obtained if one removes another data point from the fit. 

We note that bosonic string models (see
e.g. \cite{string:ambjorn,lat89:michael,string:caselle}) predict
$\Delta = 2\pi /r$ for large distances $r$. However, the fact that
\eq{e_r0delta} is almost a factor two smaller than this universal
result should not be taken as evidence that an effective bosonic
string model does not describe the QCD-string. Rather we have to
remember that the universality of string models holds only for
large~$r$. For instance, for the Nambu-Goto action, the order $1/r^3$
terms of the excitations are numerically very large at $r=r_0$
\cite{string:arvis}. In contrast, the universal $-\pi/(12r)$ term
receives smaller order $1/r^3$ corrections in that particular model, and
this term has been observed for SU(2) Yang-Mills theory
\cite{pot:r0}. We will return to an analysis of the $1/r$-term in the
potential for gauge group SU(3) in the future. Similarly to what we
find, ref. \cite{string:caselle} observes that a string model
describes the Wilson loops of the three-dimensional ${\rm Z}_2$ gauge
theory only for {\em large} loop sizes.

\subsection{The ratio $\Lmax/r_0$ \label{s_lmax}}
We now turn to discuss how to set the energy scale $\Lmax$ in the
computation of the running coupling. For motivation let us recall the
significance and the definition of $\Lmax$
\cite{alpha:SU3,schlad:rainer,reviews:leshouches,lat97:martin}.  An
important point in the quantitative and even qualitative understanding
of QCD is to connect low-energy observables like hadron masses to the
running coupling measured at high energies. If one adopts a suitable
intermediate renormalization scheme, one can first connect the
perturbative high energy region non-perturbatively to the coupling at
some low-energy scale, $q=1/\Lmax$. In fact, as a result of being able
to start deep in the perturbative region, one can compute the QCD
$\Lambda$-parameter in units of $\Lmax$ \cite{lat97:martin}. Since the
continuum limit can be taken in this calculation the result is
universal, i.e. independent of the lattice action. A second step then
involves the relation of $\Lmax$ to a low-energy scale like
$\rnod$. Our aim is to perform this second step in this section.

\subsubsection{{\SF} scheme}

A possible intermediate renormalization scheme is the {\SF} scheme
\cite{SF:LNWW}. The computation of the product $\Lmax\Lambda_{\MSbar}$
for quenched QCD will be described in detail in \cite{xxx}, and a
summary has already been given in \cite{lat97:martin}. Here we compute
$\Lmax/r_0$, which will ultimately allow to express $\Lambda_{\MSbar}$
in physical units. Note that quenched QCD is completely equivalent to the
pure gauge theory in this context, since we are considering
observables that do not involve fermion fields.

In the {\SF} scheme one considers QCD with specific Dirichlet boundary
conditions in time \cite{SF:LNWW} (at $x_0=0$ and $x_0=L$), and the
running coupling is defined in this scheme through an infinitesimal
variation of the boundary values. For most details we refer to
\cite{alpha:SU3}, but for the following we need to briefly 
discuss the structure of lattice artifacts in the running coupling. To
analyze which powers of the lattice spacing may occur, one has to list
all terms which may appear in Symanzik's effective action
\cite{impr:Sym1,SF:LNWW,impr:pap1}. In contrast to the standard 
situation (torus or thermodynamic limit), the presence of surfaces in
the {\SF} means that terms accompanied by one power of the lattice
spacing are present in general. For our particular choice of boundary
conditions \cite{alpha:SU3}, only the term
\bes
 a \int \rmd^3x \tr F_{0k} F_{0k}|_{x_0=0}
\ees
and its image at $x_0=L$ are relevant. They can be cancelled by
including one appropriately chosen term in the lattice action, whose
coefficient is denoted by $\ct$. When the proper dependence of $\ct$
on the bare coupling $g_0$ is known, cutoff effects in the
renormalized coupling (and therefore also in $\Lmax$) are reduced from
$\Oa$ to $\Oasq$. The perturbative expansion for $\ct(g_0)$ is known
up to two-loop accuracy \cite{alpha:SU3,pert:2loop,pert:2loop_prep},
\bes
 \ct = 1 - 0.089 g_0^2 - 0.030 g_0^4 + \ldots \, .
 \label{e_ctpert}
\ees
Considering that we have to insert values of $g_0^2 \approx 1$, the
perturbative series for this coefficient appears rather useful.
Nevertheless, as is always the case with perturbative expressions, it
is an art to attribute an uncertainty to it (unless additional
non-perturbative information is available). Below we will use both the
full expression \eq{e_ctpert} and its truncation at one-loop order
(i.e. $\ct= 1 - 0.089 g_0^2$) in our continuum extrapolation of
$\Lmax/\rnod$.

\subsubsection{Continuum extrapolation of $\Lmax/\rnod$}
The length scale, $\Lmax$, is defined as the value of $L$ for which
the running coupling, $\gbar(L)$ in the {\SF} scheme has the specific
value,
\bes
 \gbar^2(\Lmax)=3.48 \,.
\ees
For a definite discretization this entails that $\Lmax/a $ is a unique
function of $\beta$, which can be determined by simulations of the \SF
~\cite{alpha:SU3}. For $\ct$ to one-loop precision, numerical results
are listed in Table~3 of that reference.  We have repeated this
calculation with $\ct$ given by \eq{e_ctpert} and increased precision,
and our new results are presented in \tab{t_lmax} below.

  \begin{table}[htb] 
  \centering
  \begin{tabular}{ c c  c c c }
  \hline \\[-1.0ex]
  $L/a$  & $\beta$     & &  $L/a$ &  $\beta$   \\[1.0ex] 
  \hline \\[-1.0ex]
 4  & 5.959(2)   & &  8  &  6.476(3) \\
 5  & 6.118(2)   & & 10  &  6.654(3) \\
 6  & 6.257(3)   & & 12  &  6.799(3) \\
 7  & 6.374(3)   & & 16  &  7.026(4)
 \\[1.0ex]
  \hline 
  \end{tabular} 
\caption[t_lmax]{\footnotesize
Bare couplings vs. lattice size at $\gbar^2(L)=3.48$ and for 
two-loop $\ct$ as given in \protect\eq{e_ctpert}.
\label{t_lmax}
 }
  \end{table}

\begin{figure}[tbp]
\hspace{0cm}
\vspace{-0.0cm}

\centerline{
\psfig{file=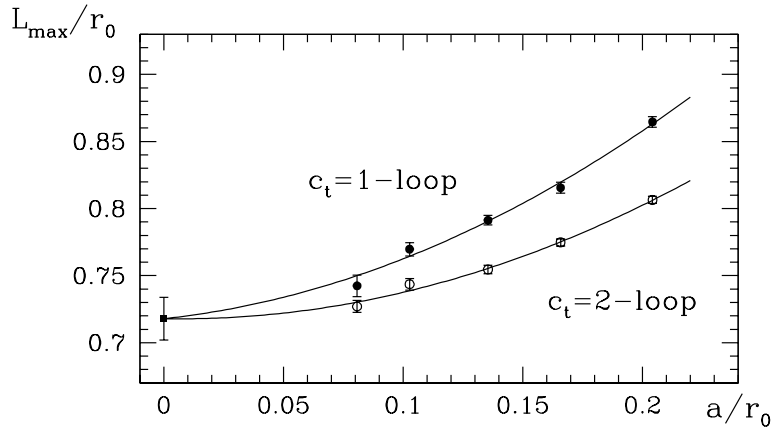,width=10cm}
}
\vspace{-0.0cm}
\caption{\footnotesize Continuum extrapolation of $\Lmax/r_0$.
         The two curves show the fit \eq{e_cont_extr_ratio} who's 
         value in the continuum limit is indicated by the square.
\label{f_lmax_r0}}
\end{figure}

In order to form the ratio $\Lmax/r_0$ we need both quantities
$\Lmax/a$ and $r_0/a$ for the same values of $\beta$. One of the data
sets has to be interpolated. We used a linear interpolation
$\beta =l_0 + l_1 \ln(\Lmax/a)$, propagating the statistical errors. This
linear interpolation is well justified: since the complete data sets
can be fitted by adding a term, $l_2 [\ln(\Lmax/a)]^2$, with a small
coefficient, $l_2$, a linear function in $\ln(\Lmax/a)$ is locally an
excellent approximation. As an alternative, we have also performed the
mentioned global fit (with $l_2$) and taken the fit as a
representation of the data. The final conclusions remain unchanged.
\footnote{
When we take $\ct$ to one-loop precision, the errors of $\Lmax/r_0$ are
dominated by the errors in $\gbar^2(\Lmax)$, while for our new data,
the errors in $r_0/a$ are the larger ones.  Therefore -- despite the
fact that the same values of $\rnod/a$ enters both data sets -- the
data points shown in \fig{f_lmax_r0} are only weakly correlated. We
neglect their correlation in the following. To be complete, we note
that in fact a second source of statistical correlations of data
points within each set is the interpolation $\beta = l_0 + l_1
\ln(\Lmax/a)$, where in most cases one simulation point contributes to
two data points in \fig{f_lmax_r0}.}

As regards the data for $\Lmax/\rnod$ shown in \fig{f_lmax_r0}, we
note that the two-loop term $-0.03\,g_0^4$ has quite a significant
effect of the order of $8\%$ at $a/r_0=0.2$, and its inclusion does
indeed reduce lattice artifacts. On the other hand, some $15\%$ of
cutoff effects remain at $a/r_0=1/5$. We conclude that the
perturbative expression \eq{e_ctpert} is very useful, but due to the
significant effect of the two-loop term we should not assume that
the remainder can be neglected.

We are led to model both data sets in a combined fit according to
\bes
  {\Lmax \over r_0} = \left.{\Lmax \over r_0} \right|_{a=0} 
                      + \rho_1^{(i)} {a \over r_0}
                      + \rho_2^{(i)} {a^2 \over r_0^2}\,, 
                      \quad i=\hbox{ 1-loop, 2-loop}\,,                
  \label{e_cont_extr_ratio}
\ees
where $\left.{\Lmax \over r_0} \right|_{a=0}$ and the 4
$\rho$'s are fit parameters. This fit yields a continuum
extrapolation
\bes
 \left. {\Lmax \over r_0}\right|_{a=0} = 0.718(16)\,.
 \label{e_lmax_r0_cont}
\ees
The figure shows nicely that the fit prefers a very small term linear
in the lattice spacing after two-loop $\Oa$ improvement. A value
consistent with our final result \eq{e_lmax_r0_cont} is also obtained
if one extrapolates just quadratically (the lower data points with
two-loop $\Oa$ improvement). However, the statistical error of such an
extrapolation, which is five times smaller than the one quoted above,
appears unrealistically small. Only with a solid, non-perturbatively
determined function $\ct(g_0)$ could we perform such a fit with
confidence.

\section{Conclusions \label{s_concl}}
In this paper we have presented a precision computation of the
low-energy reference scale $r_0$ in SU(3) Lattice Gauge Theory over a
large range of $\beta$, in which the lattice spacing varies by a
factor of four. Our results at individual values of $\beta$, which are
displayed in \tab{t_r0}, are supplemented by an interpolating
parametrization, \eq{e_poly_fit}, which provides estimates for $r_0/a$
at arbitrary values of $\beta$ in the whole range of
$5.7\leq\beta\leq6.57$.

Our study represents a significant improvement over previous
calculations through the combined effects of variance reduction by
means of the multi-hit technique, high statistics and the use of a
variational approach. Finally, a precise definition of the force has
been adhered to throughout our calculation. We observe a 2\% deviation
of our result at $\beta=5.7$ compared to ref.~\cite{pot:r0_SCRI}.

Our results for the scale $r_0$ have been applied in the continuum
extrapolation of the gap $\Delta$ of the static quark potential, and
the low-energy scale $\Lmax$, used to set the scale in the computation
of the running coupling. In the case of $\Delta$, the analysis
revealed sizeable cutoff effects of 10--15\% at lattice spacings of
$a\approx0.1\,\fm$. Nevertheless, a reliable continuum extrapolation
could be performed, showing that {\em at distances $r \approx r_0$},
the gap $\Delta$ is far from the universal value of bosonic string
theories.

The fact that $r_0/a$ has been calculated for small lattice spacings
is crucial for a stable extrapolation of $\Lmax/r_0$, since both $a$
and $a^2$ lattice artifacts turn out to be relevant for this
quantity. A reliable estimate for $\Lmax/r_0$ is of importance
in our study of the $\Lambda$-parameter, which will appear in a
forthcoming publication \cite{xxx}.

Our results for $r_0/a$ are presented in a form so that they can be
easily used in any kind of scaling analysis of hadronic quantities. A
typical example is the scaling behaviour of the vector mass, $m_{\rm
V}\,r_0$ for a fixed ratio of vector and pseudoscalar masses,
e.g. $m_{\rm V}/m_{\rm PS}=0.7$ \cite{lat97:hartmut,impr:SCRI}. Also,
the value of $m_{\rm proton}\,r_0$ in the continuum limit can be
obtained, which will serve to check how close $r_0$ is to the value
$0.5\,\fm$. 

\vspace{1cm}
This work is part of the ALPHA collaboration research programme. We
thank DESY for allocating computer time on the APE/Quadrics computers
at DESY-Zeuthen and the staff of the computer centre at Zeuthen for
their support. We are grateful to Martin L\"uscher for useful
discussions and Martin Hasenbusch and Mike Teper for reminding us of
ref.\,\cite{string:ambjorn}. We thank Tim Klassen and Gunnar Bali for
correspondence. Hartmut Wittig acknowledges the support of the
Particle Physics and Astronomy Research Council through the award of
an Advanced Fellowship.

%
   \bibliography{lattice}        
   \bibliographystyle{h-elsevier}   
\end{document}